\documentclass[a4paper,11pt]{article}
\usepackage{pos}

\renewenvironment{thebibliography}[1]{
  \begin{oldthebibliography}{#1}
    \setlength{\itemsep}{0em}
    \setlength{\parskip}{0em}
}
{
  \end{oldthebibliography}
}

\newcommand{\LEGO}{LEGO\textsuperscript{\textregistered}}

\usepackage{lineno}

\title{ALICE in Public Outreach and Bricks}
\ShortTitle{ALICE Outreach}

\author*[a,1]{Christian Klein-B\"{o}sing}

\affiliation[a]{Institut für Kernphysik, WWU Münster,\\
  Wilhelm-Klemm-Str. 9, Münster, Germany}

\emailAdd{Christian.Klein-Boesing@wwu.de}

\note{for the ALICE Collaboration}
\abstract{In the last two years various existing public outreach activities in ALICE have been adapted for online usage, this includes the well established particle physics masterclasses but also virtual visits to ALICE. Based on 
these foundations a six month online workshop was carried out in 2021 with the goal to design a \LEGO model of the ALICE detector at the LHC at the scale of a typical minifigure (ca. 1:40) and to motivate young people for a long term online collaboration in a particle physics project. 
The design stage of the model took half a year with regular online design sessions accompanied by input on 3D construction, detector technology, the physics questions of ALICE, virtual ALICE visits, and particle physics masterclasses. 
This stage provided first-hand experience on the dynamics of working on different sub-projects in a research collaboration and resulted in a model with more than 16 000 parts that was assembled by the participants during one weekend  in an in-person workshop. The experience gained during this construction has been used further by the young designers to optimise the model over the period of another year.}

\FullConference{%
  41st International Conference on High Energy physics - ICHEP2022\\
  6-13 July, 2022\\
  Bologna, Italy
}

\begin{document}
\maketitle

\begin{figure}
	\centering
	\includegraphics[width=0.9\linewidth]{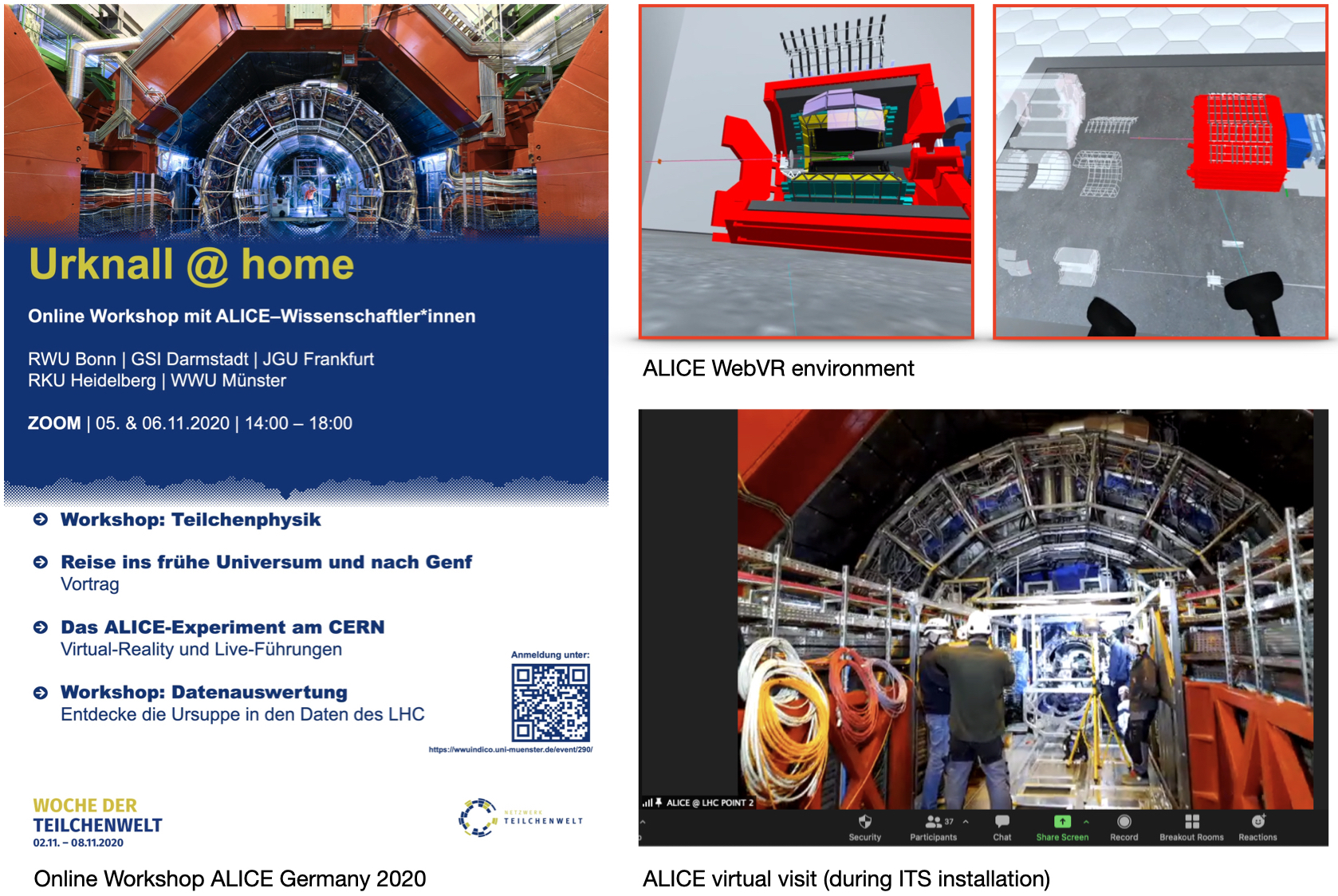}
	\caption{Advertising poster for the first virtual, nation-wide ALICE Masterclass \emph{Big Bang@home} during \emph{Woche der Teilchenwelt 2020} \cite{WocheTW},  Virtual reality implementation of the ALICE experiment in full scale view and in jigsaw mode \cite{WebVR}, ALICE virtual visit as part of virtual Masterclass}
	\label{fig_home_vr_visit}
\end{figure}
\section{Introduction}

Public outreach as a structured approach of reaching out to people with research topics, methods, technologies, and science in general serves several purposes.
It satisfies existing curiosity which is in particular present for fundamental research that addresses basic questions, such as the origin and structure of matter and the Universe.
Public outreach can help to gain new talents and enable the creation of a more diverse environment in STEM. It promotes the appreciation for fundamental research and 
educates on scientific methods and nature of science. 

Many formats of public outreach rely on the personal contact to scientists, visits to laboratories, and hands-on experiences e.g. in the particle physics masterclasses \cite{IMC}. They have been severely limited since the start of 2020 and many virtual substitutes formats have been developed.
The boosted development of online formats induced by the pandemic situation, combined with the public and in particular high school students proficiency with video conferencing and remote learning allowed to gather participants over wide areas and with a high accessibility. This has been demonstrated in first nation-wide workshops in the Netzwerk Teilchenwelt initiative \cite{Bilow:2020myc} in Germany and in various virtual international masterclasses (see Fig.\,\ref{fig_home_vr_visit}).

\section{Virtual Outreach Formats in ALICE and Prerequisites}

\begin{figure}
	\centering
	\includegraphics[width=0.9\linewidth]{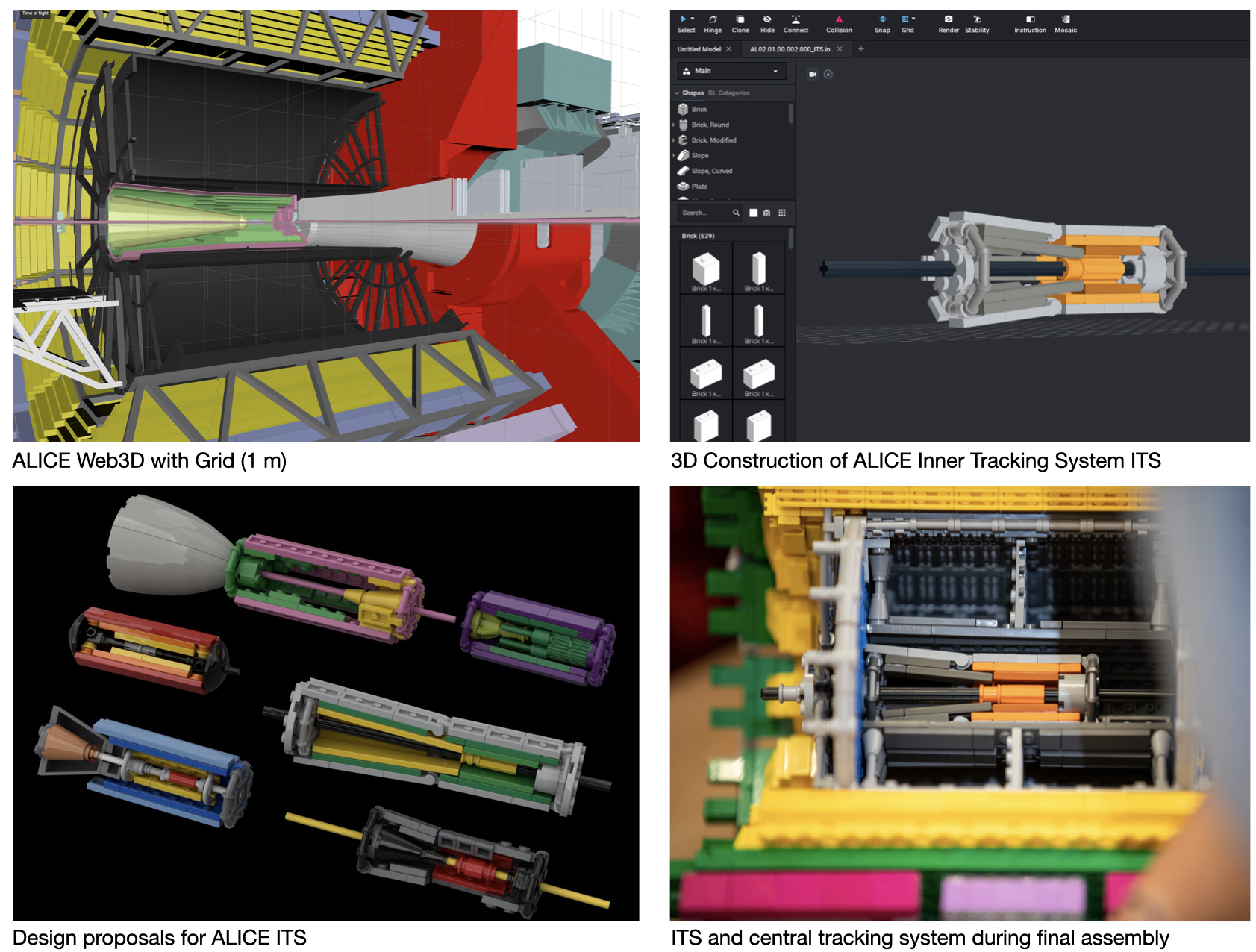}
	\caption{ALICE Web3D, Browser based exploration with optional measuring tools \cite{Web3D}, CAD program with \LEGO model of ITS \cite{studio}, Various design proposals for the ALICE Inner Tracking System and final ITS design as part of the ALICE central detectors during assembly}
	\label{fig_ITS}
\end{figure}

The particle physics masterclass program of the ALICE experiment at CERN \cite{Aamodt:2008zz} has for a long time been based on analysis of real data with a customised ROOT framework \cite{Foka:2014nqa, Averbeck}. These have been updated into web-based formats already in 2019 with a fully web-based analysis of strange particles in ALICE \cite{ALICEMC1} and of the measurement of the nuclear modification factor via python notebooks \cite{ALICEMC2}. 

At the same time the development of a 3D learning environment has been started as part of a common PhD project between the Department of Physics at the University of Münster and the
Hamburg University of Applied Sciences (HAW Hamburg). Here, the goal was to create an immersive learning experience on detector technology in particle physics on the example of the ALICE Experiment. Early on in the process it has been decided to split the development into a 3D browser application for use with PC or tablet  \cite{Web3D}  and a web-based Virtual-Reality \cite{WebVR} application.  These could be used easily in virtual formats. This has been carried out first in the context of a nation-wide week on particle physics \cite{WocheTW} and has been used successfully since then in several international masterclasses (see Fig.~\ref{fig_home_vr_visit}) \cite{WocheTW,IMC}. 

\begin{figure}
	\centering
	\includegraphics[width=0.9\linewidth]{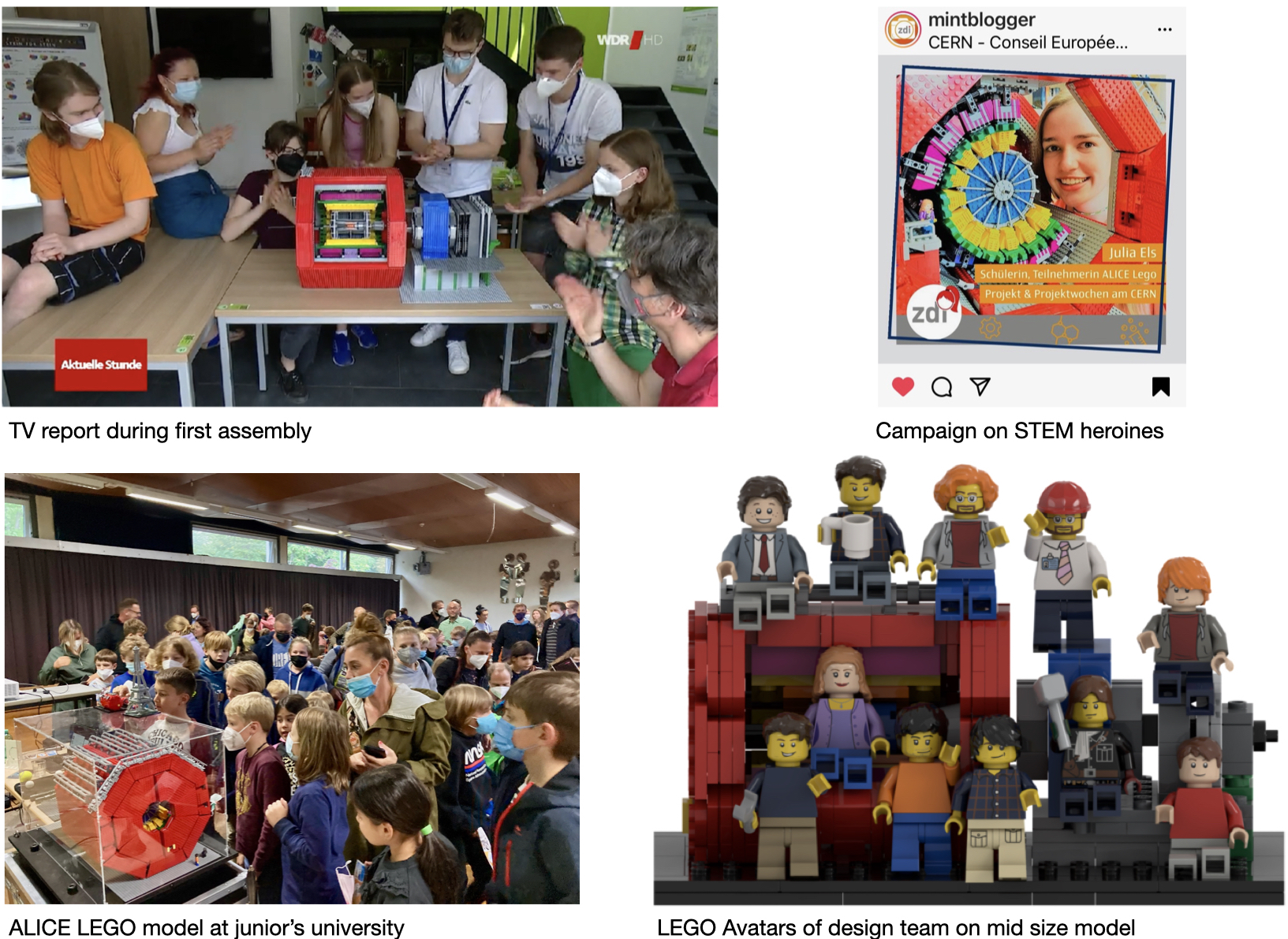}
	\caption{Echo of the first construction of the ALICE model in TV, social media and public lectures, as well as avatars of the participants with  the mid-size model.}
	\label{fig_echo}
\end{figure}

The usage of \LEGO\ as an attractive way to show LHC detectors is established since years in the project \emph{Build Your Own Particle Detector} \cite{BYOPD} by Sascha Mehlhase. Large models of the ATLAS and CMS detectors at the scale of a \LEGO\ minifigure (roughly 1:40) have already been developed and are used worldwide in outreach. The full set of LHC experiments and LHC dipoles are available as well as micro models and have attracted significant attention in the \LEGO\ IDEAS competition \cite{LHCmicro}. \section{ALICE \LEGO\ Workshop}

Inspired by the success of virtual formats and the lack of a large scale ALICE model, the outreach branch of the German ALICE community together with Netzwerk Teilchenwelt and BYOPD launched a nation-wide call to young people (aged 15 and older) for a half year workshop to construct and build an ALICE \LEGO\ model. Initially 30 participants signed up.
The design stage of the model took half a year with almost weekly meetings of 4 hours with seven tutors. This phase  included online design sessions accompanied by input on 3D construction, lectures on detector technology and structure of ALICE, the physics questions of ALICE, virtual ALICE visits, particle physics masterclasses, information on studying physics, design, and other related topics.
For the participants this stage provided first hand experience on the dynamics of working on different sub-projects in a research collaboration. The first design of the ITS has been done in a competing process (see Fig. \ref{fig_ITS}) and then subgroups have been formed for various detector systems and magnets, which had to fit together. Finally 16 participants constructed two 16 kg models with more than 16 000 parts that were assembled during one weekend  in two in-person workshops in Frankfurt and Münster. The working time was more than 100 hours in the workshop alone, plus the construction in free time. On the organisational side a significant effort has been put in the ordering process via bricklink \cite{studio} and sorting of the bricks. About 2000 Euro per model have been spent. 

The culmination in one  weekend gathered significant attention by newspapers and TV stations as shown in Fig.\,\ref{fig_echo}. Since then the final model has been a large attraction for various talk formats (e.g. for elementary school), as well as  in science camps and exhibitions. In addition also a mid-size model has been designed (see Fig.\,\ref{fig_echo})

\section{ALICE Upgrade}

\begin{figure}
	\centering
	\includegraphics[width=0.9\linewidth]{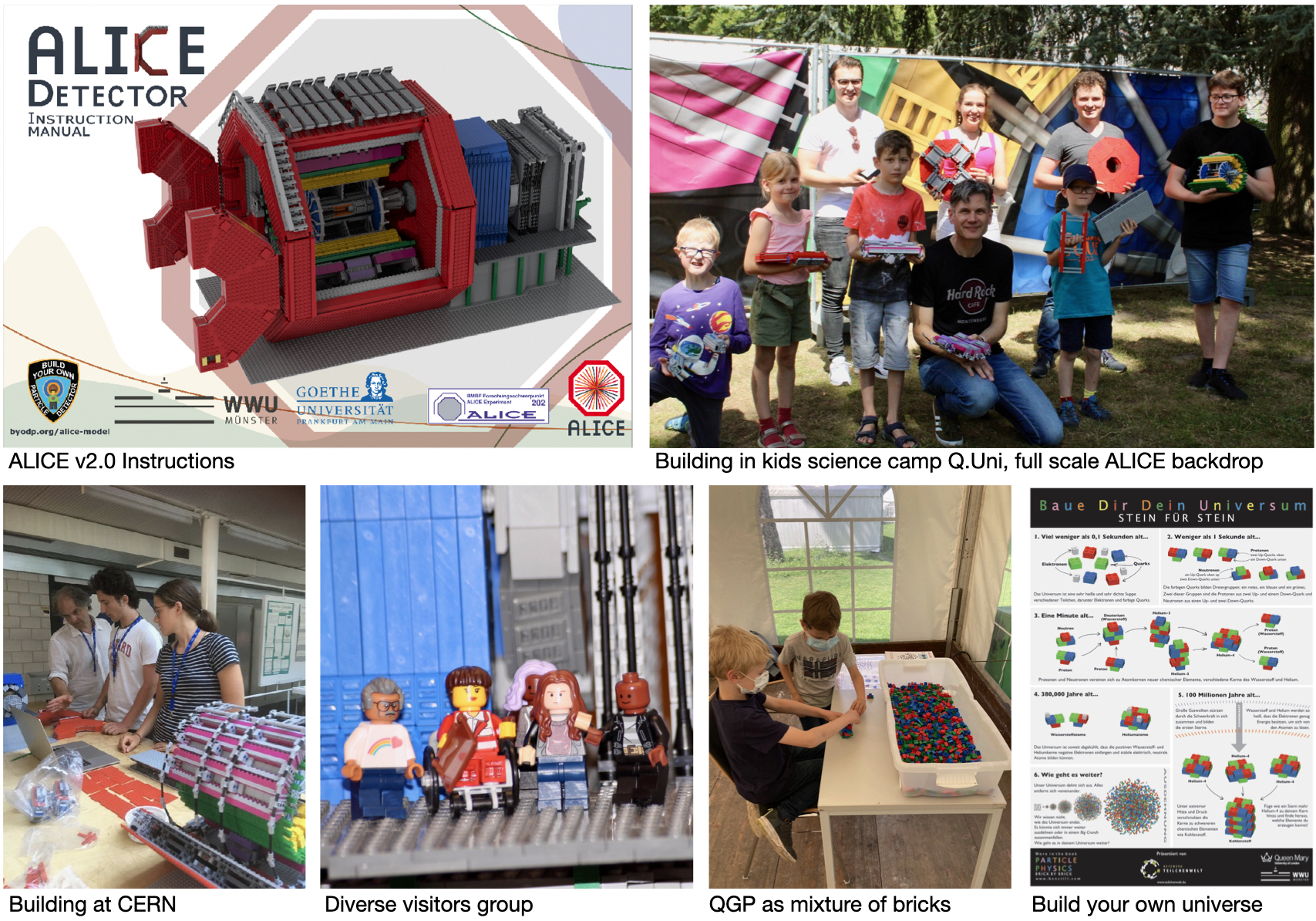}
	\caption{ALICE Instruction Cover and impressions of various events and applications of \LEGO\ in outreach and education}
	\label{fig_upgrade}
\end{figure}

The experience on mechanical stability gained during the construction with real bricks has been used  by the young designers for optimisation of the model over the period of another year. The new model (v2.0) has now more than 18 000 parts and weighs 18 kg. In parallel, detailed instructions have been developed for the full model and split into construction segments for parallel building. The instructions are complemented by information on the purpose of the various  detector components and on the ALICE experiment in general. The instructions as well as the list of parts and 3D-CAD files are available at  \cite{BYOPD}. 

Due to the detailed and segmented instructions that can be picked for different skill levels, it has been possible to build the new model with children from elementary school to high school (see Fig.\,\ref{fig_upgrade}).  The first complete model for v2.0  has been assembled by a high school class from Italy  as part of their five days visit at CERN.

\section{Outlook: Bricks as Vehicle for Outreach and Education}

Using  \LEGO\ as  a tool for science communication has been proven successful also in other areas. In the context  of a mobile exhibition (\emph{Urknall Unterwegs}, Big Bang on the Road) a coloured \LEGO\ primordial soup is used to explain nucleosynthesis in the early universe (based on the concept in \cite{Still2017}. However, the exhibit is also attractive without any explanation and serves as an ice breaker to simply start playing even for the youngest (see Fig.\,\ref{fig_upgrade}). 

The ALICE \LEGO\ model has proven to be a very attractive exhibit and eye catcher in press releases and talk announcement, since the interest in \LEGO\ covers all age groups from children to adults.
  The model in minifigure scale enables a size comparison of familiar ground and also allows one to transport messages with minifigures themselves, i.e. by using diverse groups of people around the detector
(see Fig.\,\ref{fig_upgrade}). In addition,  the simplicity in structure and the accessible color coding of the model also allowed for a rendered \emph{real size} cross section print of the \LEGO\ detector, which has been very successfully used as advertisement.

 The success of the workshop format with \LEGO\ in terms of media echo, long term commitment of young participants,  and further usage of the product can serve as a template for construction projects for other large scale particle physics experiments. This work has been supported  by BMBF (KONTAKT2 in the ErUM-Framework), Netzwerk Teilchenwelt, the DFG Research Training Group 2149 and the Joachim Herz Stiftung.


\begin{thebibliography}{99}

\bibitem{WocheTW} Woche der Teilchenwelt 2022, Nation-wide Week of Particle Physics, \url{https://www.wochederteilchenwelt.de}

\bibitem{IMC} International Particle Physics Outreach Group, \newblock {\em Particle Physics Master Classes}, \url{https://physicsmasterclasses.org}



\bibitem{Bilow:2020myc}
U.~Bilow and M.~Kobel, \emph{Netzwerk Teilchenwelt: Coordinated Outreach and Recruitment of Young Talents in Germany}, PoS ICHEP2020 (2021), 936
doi:10.22323/1.390.0936


\bibitem{Aamodt:2008zz} K.~Aamodt et~al, \newblock {\em The ALICE experiment at the CERN LHC}, \newblock { JINST}, 0803:S08002, 2008.

\bibitem{Foka:2014nqa} P.~Foka and M.~Janik. \newblock {\em ALICE Masterclass on strangeness}, \newblock {EPJ Web Conf.}, 71:00057, 2014.
\bibitem{Averbeck}  R. Averbeck et al, \newblock{\em ALICE Masterclass: The Nuclear Suppression Factor (RAA)}, \url{http://www-alice.gsi.de/masterclass/}

\bibitem{ALICEMC1} P. Nowakowski et al,  {\em ALICE Online Masterclass on strangeness}, \url{https://alice-web-masterclass.app.cern.ch/home}
\bibitem{ALICEMC2} N. Tiltmann et al,  {\em ALICE RAA Masterclass with python notebooks}, \url{https://github.com/NTW-Muenster/alice-mc-raa}


\bibitem{Web3D} P. Bhatty, S. Heusler, C. Klein-Bösing, R. Schulz-Schaeffer,  \newblock {\em ALICE Web 3D Environment}, \url{https://x3dom-alice3d.glitch.me/}
\bibitem{WebVR} P. Bhatty, S. Heusler, C. Klein-Bösing, R. Schulz-Schaeffer,  \newblock {\em ALICE Virtual-Reality Environment}, \url{https://alicewebvr.glitch.me}

\bibitem{BYOPD} S. Mehlhase,  \newblock {\em Build Your Own Particle Detector, A Particle Physics Outreach Programme}, \url{https://byopd.org/}


\bibitem{LHCmicro} N. Readiof, LHC Micro Models \href{https://ideas.lego.com/projects/5c3aec53-00d2-40a2-be73-9e2db09da86f}{LEGO ideas}, \href{https://build-your-own-particle-detector.org/models/lhc-micro-models/}{BYOPD}

\bibitem{studio} bricklink, stud.io CAD program for \LEGO\ models, \url{https://www.bricklink.com/v3/studio/download.page} 



\bibitem{Still2017} B.~Still, \newblock {\em Particle Physics Brick by Brick}, \newblock Cassel Octopus,  ISBN: {978-1844039340}, 2017.



\end{thebibliography}
\end{document}